# Empirical Measurement of Aesthetic Experience of Music

Abhishek Gupta*[1], C.M. Markan[2]
[1] (Department of Cognitive Sciences, DEI, Agra,282007, UP, India)
[2] (Department of Physics and computer Science, DEI, Agra,282007, UP, India)

*Chills or goosebumps, also called frisson, is a phenomenon that is often associated with an aesthetic experience e.g., music or some other ecstatic experience. The temporal and spatial cause of frisson in the brain has been one of the biggest mysteries of human nature. Accumulating evidence suggests that aesthetic, namely subjective, affective, and evaluative processes are at play while listening to music, hence, it is an important subjective stimulus for systematic investigation. Advances in neuroimaging and cognitive neuroscience, have given impetus to neuro-aesthetics, a novel approach to music providing a phenomenological brain-based framework for the aesthetic experience of music with the potential to open the scope for future research. In this paper, we present an affordable, wearable, easy-to-carry device to measure phenomenological goosebumps intensity on our skin with respect to real-time data using IoT devices (Raspberry pi 3, model B). To test the device subjects were asked to provide a list of songs that elicit goosebumps. Wireless earphones were provided, allowing participants to walk around and dance while listening to their music. (Some subjects moved during sessions). Results indicate that goosebumps were reliably detected by the device after visual inspection of the videos/music. The effective measurement when interfaced with neurophysiological devices such as electroencephalography (EEG) can help interpret biomarkers of ecstatic emotions. The second part of the study focuses on identifying primary brain regions involved in goosebump experience during musical stimulation.*

## I. INTRODUCTION

Music has the unique power to evoke moments of intense emotional and psychophysiological response. Called 'chills', 'thrills', 'shockers', etc., these moments are not only the subject of introspection and philosophical debate, but also of academic exploration of musical perception and perception. In the current article, we integrate existing interdisciplinary literature to define a comprehensive, testable, and ecologically valid model of transcendental psychophysiological moments in music. Music has a strong influence on people [1]. It improves memory, increases task endurance, improves mood, reduces anxiety and depression, prevents fatigue, improves pain response, and helps you exercise more effectively. Experience has three components: a subjective experience, a physiological response, and behavioral or expressive response. We experience emotions from birth and understand that something psychologically or biologically significant is affecting us. It is important to note that emotions influence not only behavior, but also objectively measurable physiological changes. An important physiological indicator [2] that has received a lot of attention is chills. Aesthetic chills can be triggered by a variety of abstract and useful stimuli, including movies, poetry, and music. Used to describe physiologically excited examples. It is also called "arousal," "stimulation," and "tremor." Under controlled conditions, this type of physiological arousal response is called hedonic or "arousal" (e.g., under laboratory conditions). Moreover, the eeriness of music is associated with dopamine production, which is associated with increased reward network activity. It is important to note that the musical chills of this project often have physiological skin correlations in the form of goosebumps (i.e., emotional bristles, visible skin hairs). [3]

Goosebumps are a nation of the pores and skin because of cold, fear, or pleasure, wherein small bumps seem on the surface because the hairs turn out to be erect, goose pimples. As you can have observed, goosebumps tend to form when you're cold. in addition, they shape when you enjoy a sturdy emotional feeling, inclusive of extreme fear, sadness, pleasure, and sexual arousal. Goosebumps might also occur during instances of bodily exertion, even for small sports, like whilst you're having a bowel movement. Goosebumps, additionally known as piloerection, arise while the tiny muscle mass is placed at the bottom of each hair follicle settlement, inflicting the hairs on our hands, legs, and torso to stand up [4]. Goosebumps are a normal response but can also imply an underlying health situation.

This article is organized as follows. After the introduction in Section I, Section II describes how we developed the Goosebumps device. Section III describes the methodology and evaluation tools used in this study.

Section IV presents preliminary performance results using different feature selection methods. Section V describes the description and impact of this document. Section VI summarizes the work and mentions future work.

## II. EXPERIMENTAL SETUP

A). Components of the device: - The basic structure of the goosebump measuring device can be seen in Fig 1. The structure is composed of an IoT (Internet of Things) device called Raspberry pi 3 Model B; a small camera fitted on the inside of the device (not shown in the figure) with white bands that help to bind the device with the subject hand [6].

The devices work on the idea of comparing and analyzing images while setting a threshold value for certain attributes. The equipment basically takes images of the subject's skin during an experiment in which the subject listens to both like and unfavorite music and our device recognizes the typical experience like goosebumps that appear as piloerections on the skin.

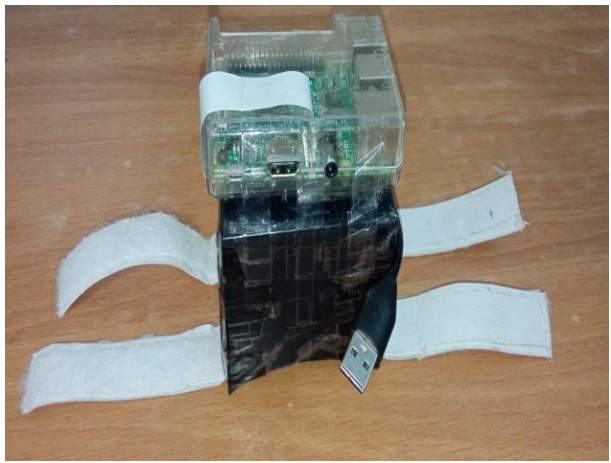

Fig.1. Representative picture of the goosebump measuring device fabricated in our lab. Parts of the device: raspberry pi settle on a black-colored stencil fitted with a camera on the inside (not shown in the picture) with white-colored straps to wear the device on hand.

B). Installation of Software Tools: - Raspbian software [7] was installed early in the process. A tool called Secure Shell (SSH) has been added for security. The camera was then connected to the Raspberry Pi and paired using the MATLAB high-level programming language with all support packages. Subsequently, similar functional codes from previous studies [8] were made available on an open-source platform and run in MATLAB for further experiments.
Furthermore, using a Raspberry Pi 3 as the base for the Goosebumps Gadget opens a whole new variety of potential for students to innovate with this device concept [8].

For example, embedded microphones, skin conductance, and heart rate sensors, among many others, can be easily attached to the Raspberry Pi and used to creatively utilize the device (for example, how is heart rate related to musical goosebumps?). To fully evaluate the device's potential, additional research is admittedly required. As we analyzed the Raspberry Pi's ability to detect the motion of hairs while goosebumps arise, the laptop showed live feeding of the camera, which shows a goosebump when the intensity of the goosebump exceeds the threshold level, making it possible to detect them [9].

## III. METHODODLOGY

We explored the idea of goosebumps in this paper and tested the results obtained from subjects. In the beginning, we ask some subjects about their favorite songs or videos that always give them goosebumps. Next, we put the device in his/her hand (Shown in Fig.2) and play the desired song while continuously monitoring the intensity of goosebumps in MATLAB. A specific intensity of goosebumps appears. After that, we were able to speak to him about some of the songs we selected for him, and they also gave us goosebumps [10].

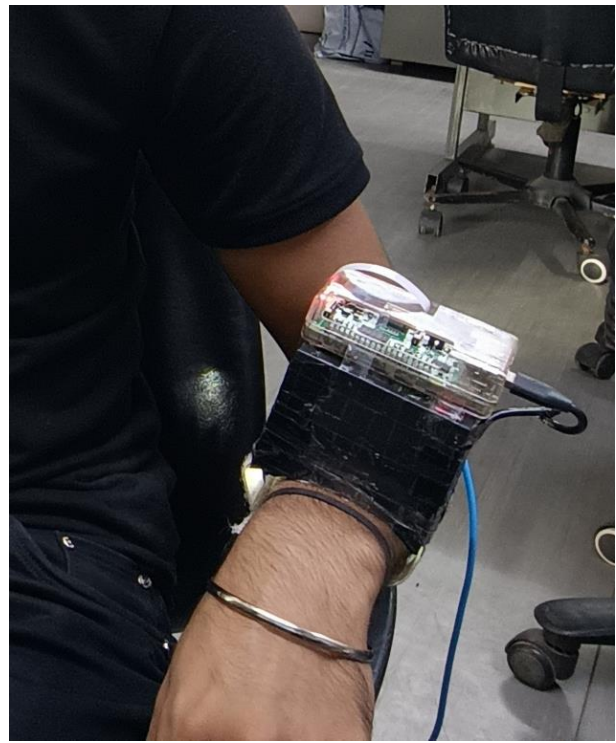

Fig 2. A participant wearing a goosebump measuring device during experimentation in the lab.

Our device captures the skin images before the stimulus and after the stimulus is given. The terrain of the skin changes during an activity like piloerection or goosebumps (Better shown in Fig.3)

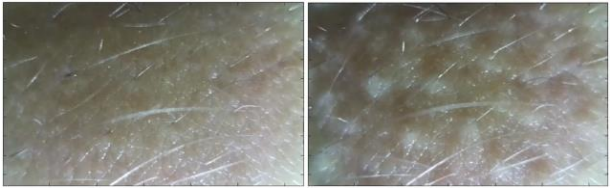

Fig. 3 Skin Image grabbed by the device (left) before (right) during goosebumps.

The electroencephalogram (EEG) is a test that analyses the electrical activity of the brain using tiny metal discs (electrodes) attached to the scalp. The next stage is to interface the device with the EEG. Even while you are asleep, electrical impulses are always being exchanged between brain cells. The primary goal of our investigation was to identify this activity (which can be seen as wavy lines in the EEG recording). The participant was given a goosebumps device to hold while holding a wet EEG hat with 64 electrodes (shown in Fig., and if goosebumps appeared, an automatic marking was made to make it easier to analyze the EEG results. (While developing a marker, we take note of the duration and severity of the goosebumps as they must be longer than the threshold) After the EEG recording is begun with various musical genres, we receive the recording that is ready for analysis. once the EEG data has been pre-processed, The EEG files are then prepared for analysis by Brainstorm using a variety of techniques, including shell comparison in the MATLAB programming language.

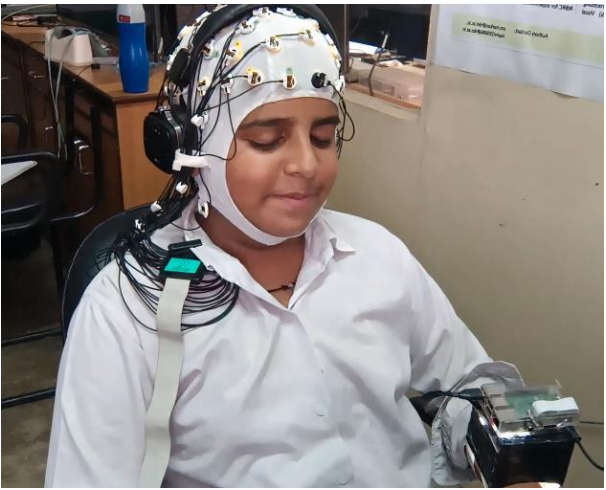

Fig.4 Subject wearing goosebumps measuring device with 64 electrodes wet EEG cap.

Our software detects goosebumps and gives us the graph between the threshold value and the time (shown in Fig.5).

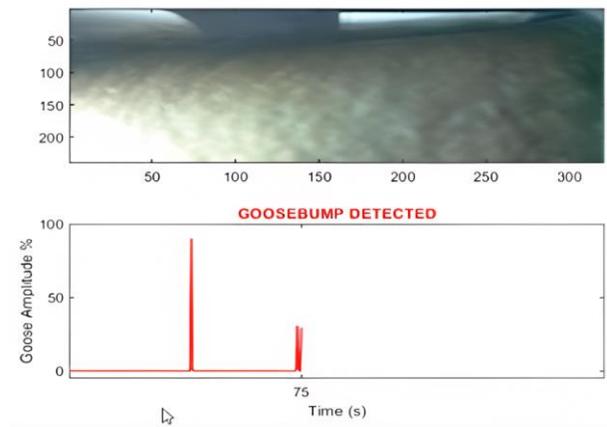

Fig. 5. Goosebumps detected in MATLAB. Graph Peak corresponds to Fig.3 (right) and flat to Fig.3 (left)

## IV. PRELIMINARY RESULTS

Firstly, the data we get from the subjects without EEG we found that the device was able to reliably detect most goosebumps experienced by the subject.

After interfacing the device with 64-channel EEG we recorded the data on 30 subjects (these subjects are from different backgrounds and different academic and personal skills) and after that we analyzed it.

By ERP (Event Related Potential) Method and by shell sphere methods we found some similar patterns in all the subjects in raw data. The Emotional Responses centre of our brain is the prefrontal cortex [11].

On preliminary group analysis, we observed some similarity of activity in most of our subjects, specifically activity in the frontal region before goosebumps (Fig 6.a.) A type of combining activity is shown before the goosebumps condition arises, During goosebumps (Fig 6.b.) A perfectly rated activity is shown in the pre-frontal cortex region of the brain, and after goosebumps (Fig 6.c.) The activity in the brain slows down that arise because of goosebumps detected by the device. This activity needs to be further tested on a larger group of subjects through EEG/MEG.

Interestingly activity profiles bear a close similarity to the perception of beauty as detected experimentally using fMRI by Semir Zeki, a professor of neuroaesthetics at University College London. According to him the mOFC, part of the reward and pleasure centers of the emotional brain, seems to light up when you're experiencing something beautiful. (Ishizu & Zeki, 2011)

Initial results of EEG after pre-processing the data using different analysis techniques discussed in the paper.

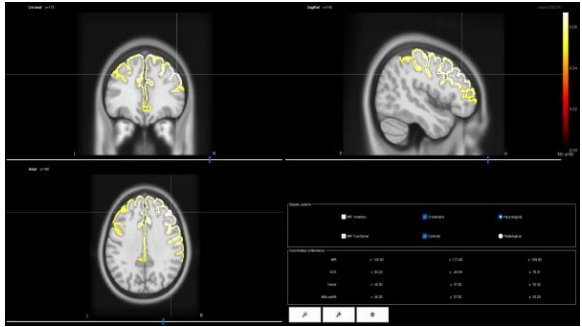

Fig. 6.a Brain activity before goosebumps

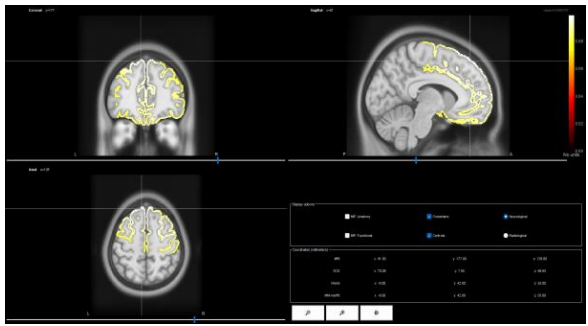

Fig. 6.b Brain activity during goosebumps

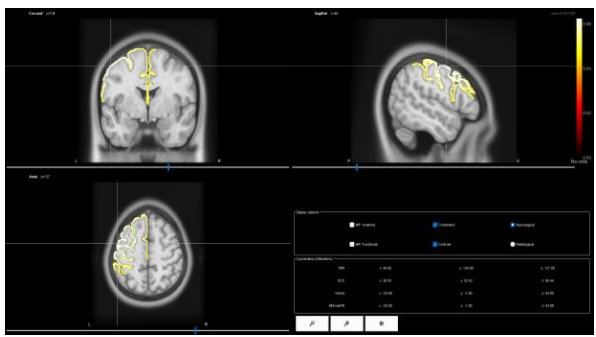

Fig. 6.c Brain activity after goosebumps

In another recent study [12], carried out using high-density EEG by Thibault Chabin a Ph.D. student at The University of Burgundy Franche-Comté in France and published in Frontiers of neuroscience also found that a specific activity over the prefrontal regions (e.g., bilateral insula, OFC, and SMA) increase in power when people get the chills from emotionally moving pieces of music.

## V. DISCUSSION AND IMPACT

There is a goosebumps device with the capability of measuring real-time goosebumps without movement affecting it.

In contrast to previously created physiological recording and biofeedback systems based just on EDA, heart rate, a mix of the two, and even EEG signals, the device's ability to examine music-evoked emotion in real-world settings is a key feature [13].

It would be possible to study the connection between emotion and music learning in the classroom since kids, teens, and adults are better able to learn and perform in groups than as individuals separated in a group context.
By monitoring the provoked emotional responses while the patients voluntarily participate in the musical interventions, a more capable and well-tested version of the Device might be utilized in this situation to evaluate the efficacy of these clinical procedures. We also believe that a more advanced model of the device may have significant applications in the music industry because it adds an additional layer of quantifiable data that can enhance our impression of the music. As a result and keeping in mind that humans are naturally sociable beings, the device's biofeedback feature might allow us to express our feelings to other concertgoers and performers while dancing along to our favorite band's song without skipping a beat [14].
With the Pi 3 may be readily attached to the device and utilized in creative ways because it has embedded microphones, skin conductance, heart rate, and other sensors. More study is necessary to gain a better comprehension of the device's potential.

## VI. CONCLUSION

The Raspberry Pi is a wearable sensor that measures goosebumps as an accurate physiological marker of musical emotions. It is necessary to refine the device and test it further in the future.

During goosebumps, we saw certain patterns of activity on the frontal areas with EEG interfacing.
Preliminary evidence suggests that the methodology is reliable when utilized by persons of diverse races and ethnicities, even though it should be resilient to variations in skin tone.

We need to validate our findings through further research.

## REFERENCES

1) McCrae, R. R. (2007). Aesthetic chills as a universal marker of openness to experience. Motivation and Emotion, 31, 5–11. https://doi.org/10.1007/s11031-007-9053-1

2) Al-Nafjan, A., Hosny, M., Al-Ohali, Y., & Al-Wabil, A. (2017). Review and Classification of Emotion Recognition Based on EEG Brain-Computer Interface System Research: A Systematic Review. Applied Sciences, 7(12), Article 12. https://doi.org/10.3390/app7121239

3) Ravaja, N., Saari, T., Salminen, M., Laarni, J., & Kallinen, K. (2006). Phasic Emotional Reactions to Video Game Events: A Psychophysiological Investigation. Media Psychology - MEDIA PSYCHOL, 8, 343–367. https://doi.org/10.1207/s1532785xmep0804_2

4) Bajaj, V., & Pachori, R. B. (2015). Detection of Human Emotions Using Features Based on the Multiwavelet Transform of EEG Signals (A. E. Hassanien & A. T. Azar, Eds.; Vol. 74, pp. 215–240). Springer International Publishing. https://doi.org/10.1007/978-3-319-10978-7_8

5) Alakus, T. B., Gonen, M., & Turkoglu, I. (2020). Database for Emotion Recognition System Based on EEG Signals and Various Computer Games—GAMEEMO. 3. https://doi.org/10.17632/b3pn4kwpmn.3

6) Ghosh, L., Saha, S., & Konar, A. (2021). Decoding emotional changes of android-gamers using a fused Type-2 fuzzy deep neural network. Computers in Human Behavior, 116, 106640. https://doi.org/10.1016/j.chb.2020.106640

7) Pelofi, C., Goldstein, M., Bevilacqua, D., McPhee, M., Abrams, E., & Ripollés, P. (2021, June 1). CHILLER: A Computer Human Interface for the Live Labeling of Emotional Responses. International Conference on New Interfaces for Musical Expression. NIME 2021. https://doi.org/10.21428/92fbeb44.5da1ca0b

8) Bevilacqua, F., Engström, H., & Backlund, P. (2019). Game-Calibrated and User-Tailored Remote Detection of Stress and Boredom in Games. Sensors, 19(13), https://doi.org/10.3390/s19132877

9) Kivikangas, J. M., Chanel, G., Cowley, B., Ekman, I., Salminen, M., Järvelä, S., & Ravaja, N. (2011). A review of the use of psychophysiological methods in game research. Journal of Gaming & Virtual Worlds, 3, 181–199. https://doi.org/10.1386/jgvw.3.3.181_1

10) Ishizu, T., & Zeki, S. (2011). Toward A Brain-Based Theory of Beauty. PLOS ONE, 6(7), e21852. https://doi.org/10.1371/journal.pone.0021852

11) Murugappan, M., Rizon, M., Nagarajan, R., Yaacob, S., Hazry, D., & Zunaidi, I. (2008). Time-Frequency Analysis of EEG Signals for Human Emotion Detection (N. A. Abu Osman, F. Ibrahim, W. A. B. Wan Abas, H. S. Abdul Rahman, & H.-N. Ting, Eds.; Vol. 21, pp. 262–265). Springer Berlin Heidelberg. https://doi.org/10.1007/978-3-540-69139-6_68

12) Chabin, T., Gabriel, D., Chansophonkul, T., Michelant, L., Joucla, C., Haffen, E., Moulin, T., Comte, A., & Pazart, L. (2020). Cortical Patterns of Pleasurable Musical Chills Revealed by High-Density EEG. Frontiers in Neuroscience, 14. https://www.frontiersin.org/articles/10.3389/fnins.2020.565815

13) Kim, J., & André, E. (2008). Emotion recognition based on physiological changes in music listening. IEEE Transactions on Pattern Analysis and Machine Intelligence, 30(12), 2067–2083. https://doi.org/10.1109/TPAMI.2008.26

14) The circumplex model of affect: An integrative approach to affective neuroscience, cognitive development, and psychopathology—PubMed. (n.d.). Retrieved February 12, 2023, from https://pubmed.ncbi.nlm.nih.gov/16262989/.